\documentclass[prb,twocolumn]{revtex4}
\usepackage{hyperref}
\usepackage{graphicx}
\usepackage{epsfig}
\usepackage{amssymb,latexsym,amsmath}

\usepackage{ulem}
\usepackage{color}

\begin{document}

\title{ \Large{ \bf Powerful Coulomb-drag thermoelectric engine} }

\author{A.-M. Dar\'{e}}\email{Anne-Marie.Dare@univ-amu.fr} 
\author{P. Lombardo}
\affiliation{ Aix Marseille Univ, Univ Toulon, CNRS, IM2NP UMR 7334, \small \it  13397, Marseille, France}

\date{\today}

\begin{abstract}
We investigate a thermoelectric nano-engine whose properties are steered by Coulomb interaction. The device 
whose design decouples charge and energy currents is 
made up of two interacting quantum dots connected to three different reservoirs. 
We show that, by tailoring the tunnel couplings, this setup can be made very attractive for energy-harvesting prospects, due to a delivered power that can be of the order of the quantum bound [R. S. Whitney, Phys. Rev. Lett. {\bf 112}, 130601 (2014); Entropy {\bf 18}, 208 (2016)], 
with a concomitant fair efficiency.
To unveil its properties beyond the sequential quantum master equation, we apply a nonequilibrium noncrossing approximation in the Keldysh 
Green's function formalism,
and a quantum master equation that includes cotunneling processes. 
Both approaches are rather qualitatively similar in a large operating regime where sequential tunneling alone fails.

\end{abstract}

\maketitle

\section{Introduction}

\begin{figure}[h!]
\begin{center}
\includegraphics[width=.48\textwidth]{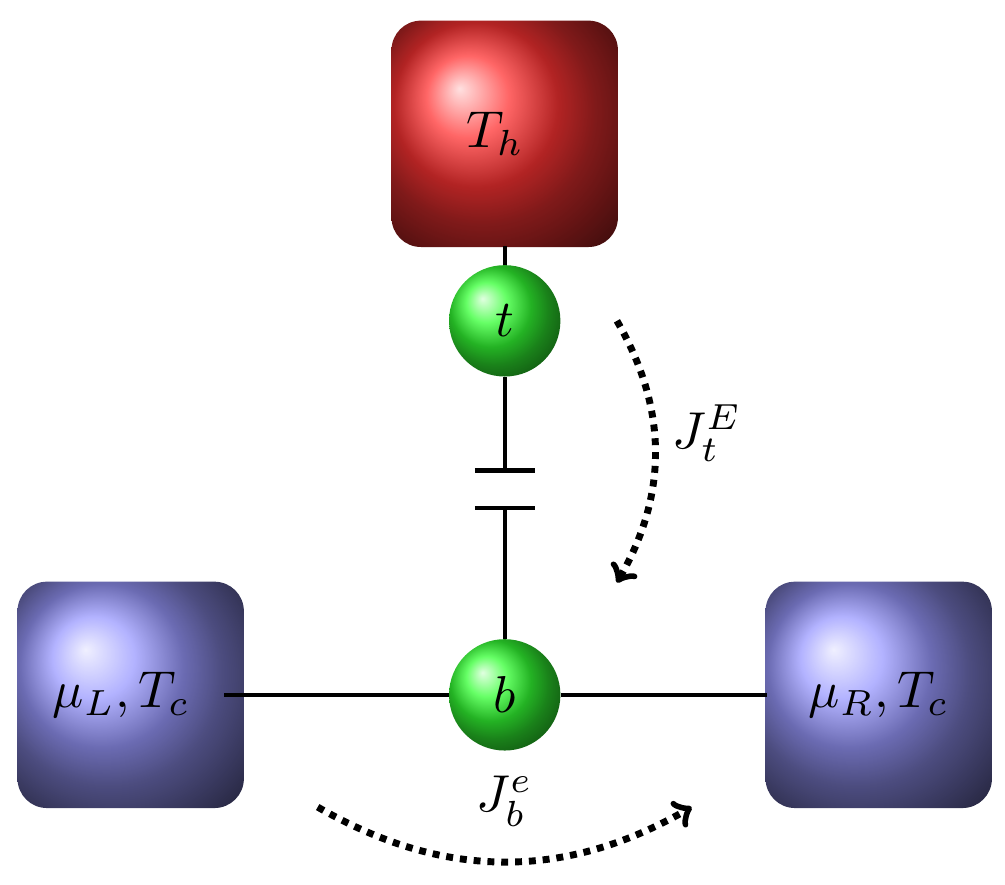}
\caption{Schematic representation of the device and of currents of interest: The top dot is connected to a hot reservoir and coupled by Coulomb interaction to a second dot (bottom). The latter is connected to two cold reservoirs through which a bias voltage can be applied. The energy current
$J^E_t$ flows from the top dot, while the electric current $J^e_b$ flows through the bottom one.}
\label{Dispo}
\end{center}
\end{figure}

A growing interest in nanodevices for energy harvesting from a temperature difference has expanded recently (see, e.g. Refs.~\onlinecite{Benenti16,Sothmann15,Zebarjadi12}). 
In addition to the prevalent two-reservoir-one-dot device, a three-terminal thermoelectric (TTTE) setup, as shown in Fig.~\ref{Dispo},
emerged a short time ago~\cite{Sanchez11}. 
One of its assets is its ability to decouple charge and heat currents. 
In this system, the working principle relies on Coulomb interaction, which triggers the output power.
The interest in such systems is not only theoretical but also experimental~\cite{Roche14,Thierschmann15-1,Thierschmann15-2,Koski15}.
Energy harvesting from voltage fluctuations in this type of device has also been established experimentally~\cite{Hartmann15}.

Other agents for energy harvesting in multiterminal systems have also been proposed~\cite{Entin-Wohlman10} based on electron-phonon or electron-photon interactions; see e.g.
Ref.~\onlinecite{Thierschmann16} and references therein.  
The present TTTE setup has been realized experimentally as an engine~\cite{Thierschmann15-1} and also as a system for thermal gating, that is, 
steering of a charge current by the temperature of a remote reservoir~\cite{Thierschmann15-2}.
It has also been investigated theoretically in the refrigerator mode~\cite{Zhang15}.
Theory for this system is not yet very advanced, the only approach being the $T$-matrix-based sequential quantum master equation (QME), which has 
its own range of validity -- weak dot-lead coupling -- and can account for the experimental results of Refs.~\onlinecite{Thierschmann15-1,Thierschmann15-2}. It can also be supplemented with full-counting statistics 
to assess current fluctuations~\cite{Nazarov,Sanchez12,Sanchez13}. 
Incidentally, Coulomb drag~\cite{Narozhny} has received much attention recently in related circuits, for which the QME up to cotunneling processes has been investigated~\cite{Jauho16,Keller16,Lim17}, however this was outside the thermoelectric context and in the absence of any temperature bias.

Our goal is to enlarge the parameter range in which the present setup was previously studied~\cite{Sothmann15}.
We apply, therefore, two radically different approaches to go beyond the sequential QME. 
The first one is an out-of-equilibrium generalization~\cite{Wingreen94,Hettler98} of the noncrossing approximation (NCA) that enables the computation of nonlinear charge and energy currents.
This technique has been previously used in a thermoelectric context, in the case of a double-orbital correlated single dot connected to two terminals~\cite{Azema12}, for which it was shown that the Kondo physics can be opportune to boost the device performance, ensuring a fair power output without eroding efficiency. 
However, we use the NCA here for rather high temperatures, for which the Kondo physics is not expected to occur.
The second approach is a QME theory that includes non local cotunneling processes. 
We show that the comparisons between the cotunneling QME and the NCA give satisfactory agreement, especially for charge-current stopping voltage, in some parameter range where sequential tunneling
alone is failing. 

The NCA allows us to 
show that, due to its topology, this TTTE device outperforms the properties of a two-terminal setup, delivering an output power 
that can be of the order of the quantum bound~\cite{Whitney14,Whitney15,Whitney16}, concomitantly with a significant fraction of the Carnot efficiency. 

The paper is organized as follows: after a presentation of the formalism and techniques (whose details are relegated to the Appendixes), the results are displayed, after which we conclude. 

\section{The model} 

\subsection{Hamiltonian and parameters}
The dots -- index $t$ or $b$ for top or bottom -- are described by a nondegenerate orbital each, and they
 are coupled by a nonlocal Coulomb repulsion $U$; this is depicted graphically in Fig.~\ref{Dispo} by a capacitive coupling.
The three reservoirs [top ($t$), left bottom ($L_b$) and right bottom ($R_b$), respectively]  are supposed to be non-interacting Fermi seas, with possibly their own chemical potentials and temperatures.
The Hamiltonian describing the present device reads $H = H_0+H_T$, where the disconnected part for dots and leads is, in obvious notations,
\begin{equation}
H_0=  \epsilon_t \hat{n}_t+ \epsilon_b \hat{n}_b+ U \hat{n}_t \hat{n}_b +\sum_ {\alpha=t,L_b,R_b}H_{0\alpha }  \ ,
\label{h0}
\end{equation}
with $H_{0\alpha }= \sum_{k} \epsilon_{k \alpha}\hat{n}_{k \alpha}$. Hybridization between dots and leads reads
\begin{equation}
H_T =\sum_{k} \Bigl( V_{k t} c^\dagger_{k  t} d_t +H.c. \Bigr)
+\sum_{\beta_b=L_b, R_b}\sum_{k} \Bigl( V_{k \beta_b} c^\dagger_{k \beta_b} d_b +H.c. \Bigr) .
\label{ht}
\end{equation}
The hybridization parameters will depend on the wave vector only through energy: $V_{k \alpha}=V_\alpha(\epsilon_{k \alpha})$~\cite{JauhoWingreenMeir1994}.
They will enter the present calculations through $\Gamma_\alpha(\epsilon)=2 \pi \rho_{\alpha}(\epsilon) |V_\alpha(\epsilon) |^2$ with 
$\rho_\alpha(\epsilon)$ the $\alpha$-lead density of states.

The three-terminal device performance is driven by charge fluctuations of the two dots~\cite{Sanchez11}.
The nondegenerate-orbital occupancy fluctuates, and for fermions the fluctuations are maximized at half-filling due to  
$\langle n^2\rangle -\langle n\rangle^2= \langle n\rangle(1-\langle n\rangle)$.
To promote this favorable situation, we choose henceforth for the two dots the following energies: $\epsilon_t =\epsilon_b\equiv\epsilon_d= -U/2$, which guarantees half-filled dots. 
Without loss of generality, the chemical potential of the top lead will be $\mu_t=0$. $V$, the voltage bias across the bottom part of the device will be applied symmetrically, with $\mu_L = - \frac{eV}{2}$ and $\mu_R = \frac{eV}{2}$. The temperature $T_t=T_h$ of the hot-top reservoir will be supposed higher than the temperature of the cold lower ones: $T_{L_b}=T_{R_b}=T_c$. 
In the case of half-filling and symmetrically applied bias, we have the following equality between Fermi functions of the two cold reservoirs:
$f_{R_b}(\epsilon_d+U) =1-f_{L_b}(\epsilon_d)$. 

To be run as a thermoelectric engine~\cite{Sanchez11}, one needs nonuniform and dissymmetrical hybridizations between the $b$-dot and its reservoirs. 
We choose 
\begin{align}
\Gamma_{t}(\epsilon) &=  \Gamma_t \ ,\nonumber \\
\Gamma_{R_b}(\epsilon) &=\Gamma_b\  \theta(\epsilon) \ \kappa(\epsilon)\ , \nonumber\\
\Gamma_{L_b}(\epsilon) &= \Gamma_{R_b}(-\epsilon) \ ,
\label{nosGamma}
\end{align}
where the function $\kappa(\epsilon)$ will be specified later. $\theta(\epsilon)$ is the Heaviside function.

\subsection{Currents in the Keldysh formalism and the NCA }
In the Keldysh formalism, stationary charge~\cite{JauhoWingreenMeir1994} and energy currents flowing outside an $\alpha$-terminal into a $d$-dot can be expressed as
\begin{eqnarray}
 \left( \begin{array}{c} J^e_\alpha \\ J^E_\alpha \end{array} \right)
& & = 
\frac{i }{\hbar} \int  \frac{d \epsilon}{2 \pi}
\left( \begin{array}{c} e \\ \epsilon \end{array} \right)
 \Gamma_\alpha(\epsilon)  \nonumber \\
& &\times \biggl[  f_\alpha (\epsilon)
G_d^>(\epsilon) + (1- f_\alpha (\epsilon)) G_d^<(\epsilon)   \biggr] \ ,
\end{eqnarray}
where $G_d^\lessgtr(\epsilon)$ are the lesser and greater $d$-dot Green's functions, $f_\alpha (\epsilon)$ is the Fermi function of the $\alpha$-reservoir, 
$e>0$ is the elementary charge, and  hybridization parameters 
$\Gamma_\alpha(\epsilon)$  were previously defined.
In the preceding  integrands, the two terms can be interpreted as a balance between in and out transfers: indeed, for fermions $i G_d^>(\epsilon) \ge 0$, whereas 
$i G_d^<(\epsilon) \le 0$~\cite{Haug}. 

We define the electric current flowing through the bottom dot from its symmetric expression $J^e_b \equiv J^e_{L_b}=-J^e_{R_b}=(J^e_{L_b}-J^e_{R_b})/2$, which leads to
\begin{align}
J^e_b =&\frac{i e}{2\hbar} \int  \frac{d \epsilon}{2 \pi}  \biggl[   G_b^>(\epsilon)
\Bigl(\Gamma_{L_b}(\epsilon) f_{L_b}(\epsilon) - \Gamma_{R_b}(\epsilon) f_{R_b}(\epsilon)\Bigr) \nonumber \\
+
& G_b^<(\epsilon)
\Bigl( \Gamma_{L_b}(\epsilon) (1-f_{L_b}(\epsilon)) - \Gamma_{R_b}(\epsilon) (1-f_{R_b}(\epsilon))\Bigr)  \biggr] \ .
\label{courantelec}
\end{align}
In the thermoelectric engine regime, for properly chosen nonproportional hybridization functions $\Gamma_{\beta_b}(\epsilon)$,  the above current can be finite even if the Fermi functions $f_{L_b}(\epsilon)$ and $f_{R_b}(\epsilon)$ are equal, that is, in the absence of any bias applied to the lower part of the device. 
In the stationary regime, the electric current flowing into the top dot must vanish,
\begin{equation}
J_t^e= \frac{i e}{\hbar} \int \frac{d \epsilon}{2 \pi} \Gamma_t(\epsilon) \Bigl[ f_t(\epsilon) G_t^>(\epsilon) 
+(1-f_t(\epsilon)) G^<_t(\epsilon)  \Bigr] = 0 \ ,
\label{courantelecnul}
\end{equation}
while the energy current flowing from the top dot to the bottom one reads 
\begin{equation}
J_t^E= \frac{i }{\hbar} \int \frac{d \epsilon}{2 \pi} \epsilon \Gamma_t(\epsilon) \Bigl[ f_t(\epsilon) G^>_t(\epsilon) 
+(1-f_t(\epsilon)) G^<_t(\epsilon)  \Bigr] \ .
\label{courantenerg}
\end{equation}
The thermal engine efficiency is defined by the ratio
\begin{equation}
\eta = \frac{\mathcal{P}}{J^E_t}=\frac{J^e_b V}{J^E_t} \ ,
\end{equation}
with $\mathcal{P}$ the output power.

As previously mentioned in Eq.~(\ref{nosGamma}), $\Gamma_{L_b}(\epsilon)$ and $\Gamma_{R_b}(\epsilon)$, are not proportional, hence one needs to evaluate two Green's functions for each dot. Indeed, the usual simplification~\cite{JauhoWingreenMeir1994} that allows to calculate only the difference $i(G^>(\epsilon)-G^<(\epsilon))$ no longer applies. 

Equations~(\ref{courantelecnul}) and (\ref{courantenerg}) look very similar, but in the present study the first cancels while the second is expected to 
be the thermal energy supply. This constitutes a key constraint for relevant approaches to study the problem at hand. For example, the many-body Hubbard-I approximation~\cite{HubbardI,Hewson66} is too simple: It can be shown that within this method, the term in square brackets in Eq.~(\ref{courantelecnul}) vanishes, and as a consequence the energy current [Eq.~(\ref{courantenerg})] too, due to an inadequate treatment of fluctuations.
The Ng method~\cite{Ng96}, which can handle Kondo physics~\cite{Dong02}, would also be inefficient to account for the present device properties. 

This led us to develop a variant of the NCA for the present system. In this approach, the Green's functions of the two fermions residing, respectively, on the top and bottom dots, are expressed in terms of fictitious particle Green's functions~\cite{Haule01}. 
The approximation of the NCA essentially lies in the self-energy choice for these four fictitious particles, two fermions and two bosons. The details of these self-energies, as well as the two real-fermion Green's function expression, are given in Appendix A. 

The NCA is a perturbative approach that presumes $ U > \Gamma$; however, contrary to the QME, there is no restriction on temperatures, except at very low temperature.
We use the finite $U$ version of the NCA, which has been shown, in its equilibrium version, to have some drawbacks~\cite{Sposetti16}, especially in the particle-hole symmetric case, which is the situation at hand. 
If so, the two fictitious fermions are degenerate, as well as the two fictitious bosons. As a result, the two real fermions, residing on the top and bottom dots, are described by the same Green's functions, notwithstanding the fact that the surroundings of the two dots may be quite different. We argue that this caveat is not prohibitive: first of all, this overstated symmetry exists also in the QME, when including cotunneling processes, as proved even for $\Gamma_t \neq \Gamma_b$.  Second, in the following we consider rather high temperatures, for which the problem is less acute.
 
As discussed in Refs.~\onlinecite{Sposetti16,Tosi11}, the easier way to cure this symmetry drawback would be to include vertex corrections in the fictitious-particle self-energies, an approximation called the one crossing approximation (OCA). If it was undertaken at equilibrium~\cite{Sposetti16}, it would be 
a formidable task to adapt this idea in the present out-of-equilibrium case for many reasons:
First, in the present calculations, the lesser and greater self-energies of the four fictitious particles are needed, and not only the retarded ones. Second, at the OCA level the self-energies are 
products of five Green's functions. Hence it would be very cumbersome to apply the Langreth rules, which are needed for out-of-equilibrium Green's functions.
Last but not least, in the NCA or the OCA, the interdependent Green's functions are obtained from self-consistent numerical evaluations. Accessing reliable results for eight coupled functions would probably be an elusive goal. 
Finally, although we are in a different configuration, with dissymmetrical lead couplings, the temperatures investigated in the present paper are rather 
high~\cite{noteTemp}, 
higher than those below which a qualitative difference was found in the conductance evaluation in the NCA and the OCA for the two-terminal device~\cite{Tosi11}.

\subsection{$T$-matrix QME}
Up to now, to the best of our knowledge, only the sequential QME has been investigated for the present 
device~\cite{Sothmann15,Thierschmann15-1,Thierschmann15-2,Thierschmann16}. The $T$-matrix cotunneling has already been worked out, but in Coulomb drag systems without temperature bias~\cite{Jauho16,Keller16,Lim17}. Cotunneling has also been developed in the presence of a temperature bias~\cite{Gergs15}, 
but another method, namely a real-time diagrammatic method, was used to obtain the transition rates.

In the QME, the reservoir degrees of freedom are traced out, hence one focuses on the four two-dot states, noted  in an self-explanatory manner, from empty to full:  $|0\rangle$, $|t\rangle$, $|b\rangle$, and $|2\rangle$. The corresponding probabilities $P_n$ for $n=0,t,b,2$, are governed by transfer rates $\gamma_{mn}$ from state $m$ to state $n$. With the weak-coupling assumption,
the rates for tunneling-induced transition can be obtained by the generalized Fermi golden rule~\cite{Bruus}
\begin{equation}
\gamma_{mn}  = \frac{ 2 \pi}{\hbar} \sum_{i' f'} W_{i'} |\langle f' n|T|i' m\rangle |^2 \delta (E_f -E_i)  \ , 
\label{gamma}
\end{equation}
where 
$|i\rangle=|m\rangle \otimes| i'\rangle $ is the initial product state for dots and reservoirs, $|f\rangle=|n\rangle \otimes | f'\rangle$ is the final one, and $E_i$ and $E_f$ are their respective energy. 
More precisely, 
$|n\rangle (|m\rangle)$ is the double-dot state, and the corresponding energy is $\epsilon_n (\epsilon_m)$.
The states $|i'\rangle$ and $|f'\rangle$ are shorthand notations for tensorial products of three reservoir states. 
 $W_{i'}$ is the probability of the state $|i'\rangle$, assuming equilibrium for each reservoir, i.e., it is given by the grand-canonical Gibbs distribution with appropriate 
 temperatures and chemical potentials. 
 Up to the second order in $H_T$, the $T$-matrix can be written~\cite{Bruus} 
 \begin{equation}
T = H_T + H_T \frac{1}{E_i-H_0+i \eta} H_T \ ,
\label{operateurT}
\end{equation}
where $H_0$ and $H_T$ were previously defined in Eqs.~(\ref{h0}) and (\ref{ht}). 
The first term in the $T$-matrix expression represents the sequential tunneling processes, while the second term includes the cotunneling ones. 
The nature of the $T$-matrix QME requires that the energies of the double dot states are discrete, implying $\Gamma \ll T, U$, to be relevant.

In the stationary regime, the probabilities $P_n$ are obtained from the following system:
\begin{align}
\frac{d P_0}{dt} & = -(\gamma_{0t}+\gamma_{0b}+\tilde{\gamma}_{02})P_0+ \gamma_{t0}P_t+\gamma_{b0}P_b+\tilde{\gamma}_{20}P_2=0 \nonumber\\
\frac{d P_t}{dt} & = \gamma_{0t} P_0-(\gamma_{t0}+\tilde{\gamma}_{tb}+\gamma_{t2})P_t+ \tilde{\gamma}_{bt}P_b+\gamma_{2t}P_2=0 \nonumber\\
\frac{d P_b}{dt} & = \gamma_{0b} P_0+\tilde{\gamma}_{tb}P_t-(\gamma_{b0}+\tilde{\gamma}_{bt}+\gamma_{b2})P_b+ \gamma_{2b}P_2 =0\nonumber\\
\frac{d P_2}{dt} & = \tilde{\gamma}_{02} P_0+ \gamma_{t2}P_t+\gamma_{b2}P_b -(\tilde{\gamma}_{20}+\gamma_{2t}+\gamma_{2b})P_2=0 \ ,
\end{align}
where the notation highlights the distinction between cotunneling terms ($\tilde{\gamma}_{mn}$) and sequential ones ($\gamma_{mn}$).
In the particle-hole symmetric case, $\epsilon_t=\epsilon_b \equiv \epsilon_d=-U/2$, simplifications occur as detailed in Appendix B. One then gets $P_t=P_b$ and $P_0=P_2$, leading by normalization to $P_0+P_t = \frac 1 2$. Furthermore, using 
$\Gamma_t(\epsilon) =  \Gamma_t$ and $\Gamma_{L_b}(\epsilon)=\theta(-\epsilon)\Gamma_b=\Gamma_{R_b}(-\epsilon)$, we obtain
\begin{equation}
P_t = \frac \hbar 2 \frac{\gamma_{0t}+\gamma_{0b}}{\Gamma_t+\Gamma_b} \ ,
\end{equation} 
indicating that the probabilities $P_i$ are not affected by cotunneling processes for the present parameters. 
It stems from the particle-hole symmetry, indeed, as detailed in Appendix B, 
$\tilde{\gamma}_{02}=\tilde{\gamma}_{20}=\tilde{\gamma}_{tb}=\tilde{\gamma}_{bt}$, thus the cotunneling does not affect the dot occupancy. However, 
it will modify the currents. 
One can note that the dots are half-filled as expected, since $\langle n_t\rangle =\langle n_b\rangle=P_t +P_2=1/2$.

To express the electric current of interest, a more detailed version of transfer rates is needed with the specification of relevant reservoirs. 
Sequential terms imply only one reservoir,  and the superscript in $\gamma_{nm}^\alpha$ indicates the concerned one. For the drag current, cotunneling processes involve two reservoirs, one of which is always the $t$ one; thus only the second reservoir needs to be specified: For example, $\tilde{\gamma}^{R_b}_{bt}$ is the transfer rate from the $|b\rangle$ state to the $|t\rangle$ one, with the fermion on the bottom dot hopping to the $R_b$-reservoir.
The detailed expressions for all transfers in the present particle-hole symmetric model, as well as the regularization scheme, are postponed to Appendix B. 

In a general case, one should have also to consider transfer rates of the form $\tilde{\gamma}_{nn}^{\alpha \beta}$ reflecting processes in which the two-dot states do not change, however two electron hops from and to different reservoirs are implied in the process.
These terms do not enter the $P_i$ expressions but would enter the currents for some of them. They are intradot cotunneling processes, and they cancel  or do not contribute to the currents in our case, because of the present structureless dots, 
and because of our choice for hybridization functions $\Gamma_\alpha(\epsilon)$.
The only cotunneling terms implied in our calculations are delocalized processes implying the two dots, as in the Ref.~\onlinecite{Jauho16}.

Finally, the charge currents involving the bottom dot read
\begin{eqnarray}
J^e_{Lb}&=e \Bigl[&( \gamma^{L_b}_{0b} +\tilde{\gamma}^{L_b}_{02})P_0 + ( \gamma^{L_b}_{t2}+ \tilde{\gamma}^{L_b}_{tb}) P_t  \nonumber\\
& &-( \gamma^{L_b}_{b0}+\tilde{\gamma}^{L_b}_{bt}) P_b -(\gamma^{L_b}_{2t}+\tilde{\gamma}^{L_b}_{20}) P_2 \Bigr]\nonumber\\
J^e_{Rb}&=e  \Bigl[&( \gamma^{R_b}_{0b}+\tilde{\gamma}^{R_b}_{02}) P_0 + ( \gamma^{R_b}_{t2}+ \tilde{\gamma}^{R_b}_{tb}) P_t \nonumber\\
& & -( \gamma^{R_b}_{b0}+\tilde{\gamma}^{R_b}_{bt}) P_b -(\gamma^{R_b}_{2t} +\tilde{\gamma}^{R_b}_{20})P_2  \Bigr] \ .
\label{courantelec2}
\end{eqnarray}
In the stationary regime, $J^e_b \equiv J^e_{Lb}=-J^e_{Rb}$.
Gathering the results of Appendix B, we get
\begin{equation}
J^e_b = \frac{e \Gamma_b}{\hbar}\frac{1}{2(1+\Gamma_b/\Gamma_t)} \Bigl(f_{L_b}(\epsilon_d)-f_{t}(\epsilon_d)    \Bigr) + \frac e 2 \Delta \ ,
\label{courantQME}
\end{equation}
where the first right-side term is the sequential current, whereas the second one is a 
purely cotunneling contribution whose detailed expression is given in the Appendix B. 

For the present parameters [special choice of $\Gamma_{L_b}(\epsilon)$, $\Gamma_{R_b}(\epsilon)$, $\epsilon_d$, and $U$], we can evaluate the energy current supplied by the $t$-reservoir without additional calculations. This is not true in the general case in which evaluating energy or heat current is not a simple issue~\cite{Gergs15}. 
First in the $T$-matrix QME, the dot spectral functions are $\delta$-functions, which are located at $\epsilon_d$ (negative in our calculations) and $\epsilon_d+U$ 
(positive here), and there is neither broadening nor shift~\cite{Kaasbjerg15}, 
unlike in some other QME approaches; see, e.g. Ref.~\onlinecite{Gergs15}.
Second, $\Gamma_{L_b}(\epsilon)$ cancels for positive energy, such that carriers entering the bottom dot from the left reservoir can only have an energy $\epsilon_d$, such that
$J^E_{L_b}=\frac {\epsilon_d}{e} J^e_{L_b}$. 
On the other hand, $\Gamma_{R_b}(\epsilon)$ is nonzero only for positive energy, leading to 
$J^E_{R_b}=\frac {\epsilon_d+U}{e} J^e_{R_b}$.
By energy conservation in the stationary regime, we have $J^E_t+J^E_{L_b}+J^E_{R_b}=0$. Using $J^e_{L_b}=-J^e_{R_b}$, the so-called tight-coupling relation emerges:
\begin{equation}
J_t^E = \frac U e J^e_b \ ,
\end{equation}
which leads to an engine efficiency 
\begin{equation}
\eta_{\mathrm{QME}} = \frac{\mathcal{P}}{J^E_t} = \frac{eV}{U} 
\end{equation}
which was found in the sequential regime~\cite{Sothmann15}, and it persists at the cotunneling level for the present parameters.

\section{Results}

\begin{figure}[h!]
\begin{center}
\includegraphics[width=.48\textwidth]{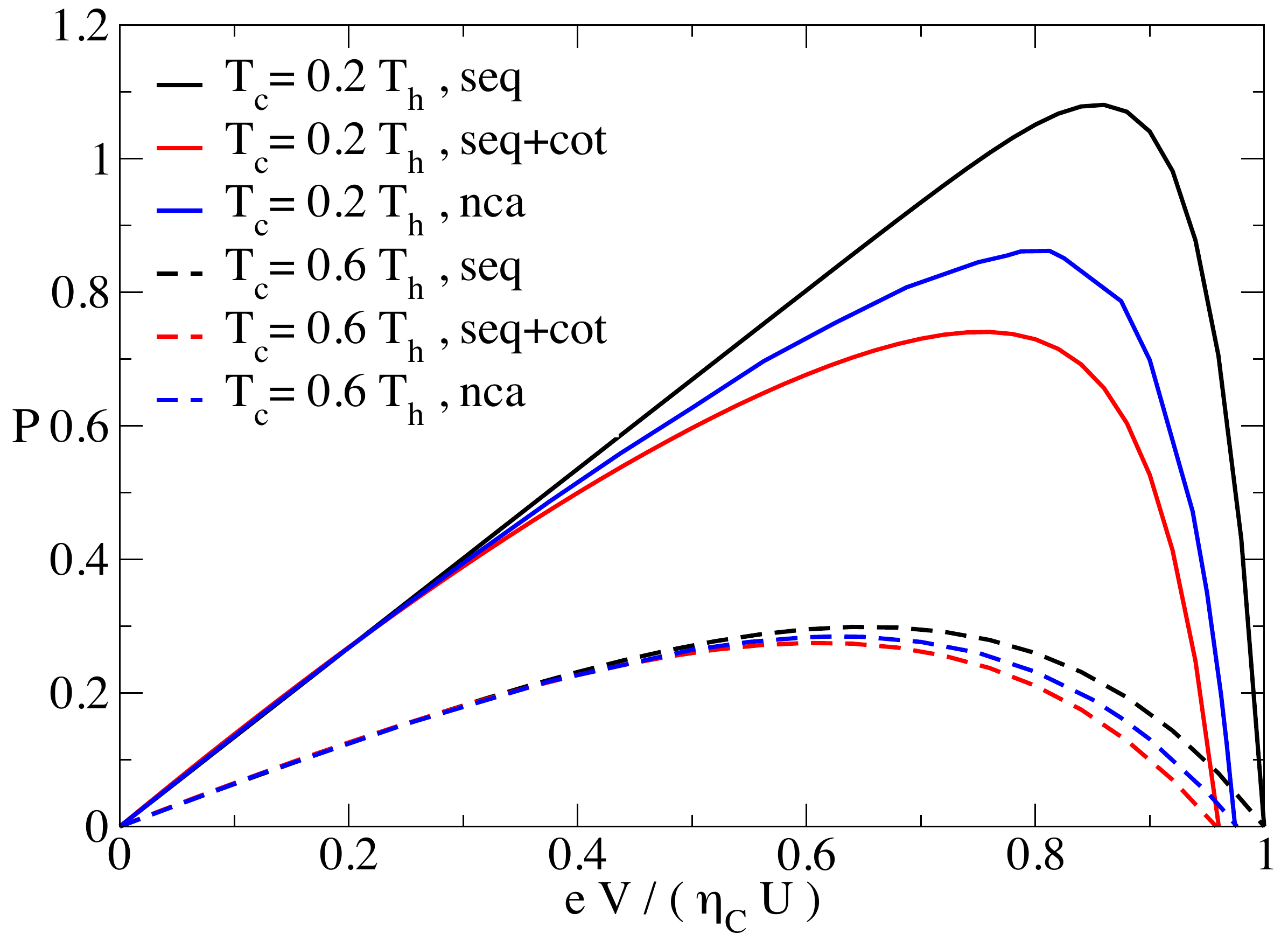}
\caption{Comparison between the NCA and the QME with (seq+cot) and without (seq) cotunneling processes: The output power $\mathcal{P}$, in $\Gamma^2/\hbar$ units, is displayed as a function of bias voltage, for $U=10 T_h$, $\epsilon_d=-U/2$, $\Gamma=T_h/100$, $T_c=0.2 T_h$ (solid lines) or $T_c=0.6 T_h$ (dashed lines). }
\label{fig1}
\end{center}
\end{figure}

We begin by quantitative comparisons between the NCA and the QME. 
We take first $\kappa(\epsilon)= 1$ in the hybridization functions [Eqs.~(\ref{nosGamma})].
Furthermore, we choose the following equality: $\Gamma_t = \Gamma_b \equiv \Gamma$. 
In Fig.~\ref{fig1} the output power is displayed for parameters for which the QME is supposed to be relevant~\cite{Sothmann15}.
Noting the Carnot efficiency $\eta_C = 1 - T_c/T_h$, the sequential current vanishes for $eV=\eta_C U$, as can be readily seen by looking at the first right-side term in Eq.~(\ref{courantQME}).
For high bias $ e V \gtrsim \frac  1 2 \eta_C U$, taking into account cotunneling processes matters, especially for low $T_c$ as $T_c = 20 \Gamma$. The NCA results are closer to these predictions, 
particularly concerning the stopping voltage, which corresponds to the bias voltage needed to cancel the current. 
As in the QME, for the present parameters in the NCA, the tight-coupling relation between energy and charge currents is nearly fulfilled, leading to an efficiency $\eta_{\mathrm{NCA}} \simeq e V/ U$.

\begin{figure}[h!]
\begin{minipage}[t]{.48\textwidth}
        \begin{center}
            \includegraphics[width=\textwidth]{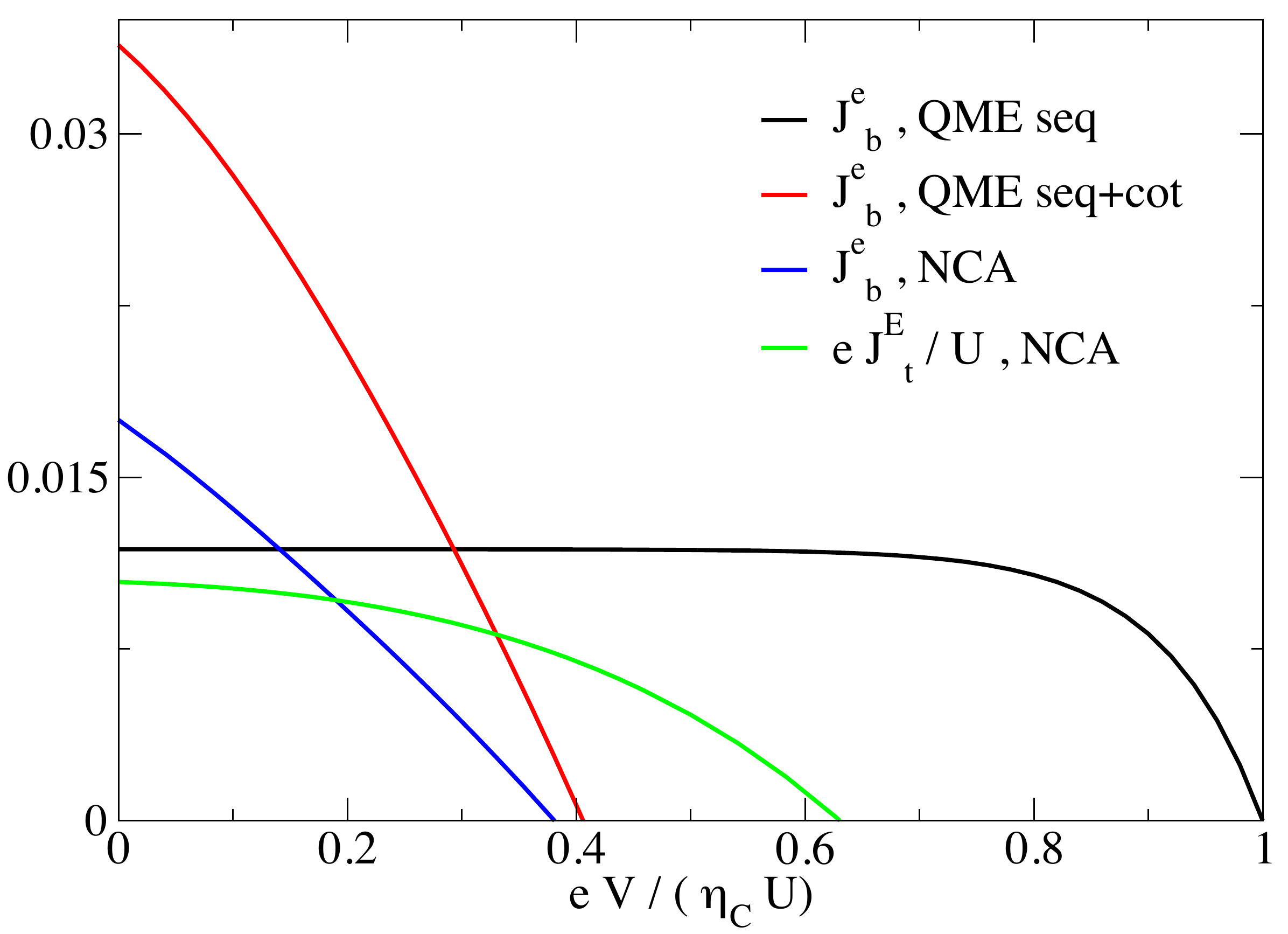}
        \end{center}
    \end{minipage}
    \begin{minipage}[t]{.48\textwidth}
            ~~\includegraphics[width=0.96\textwidth]{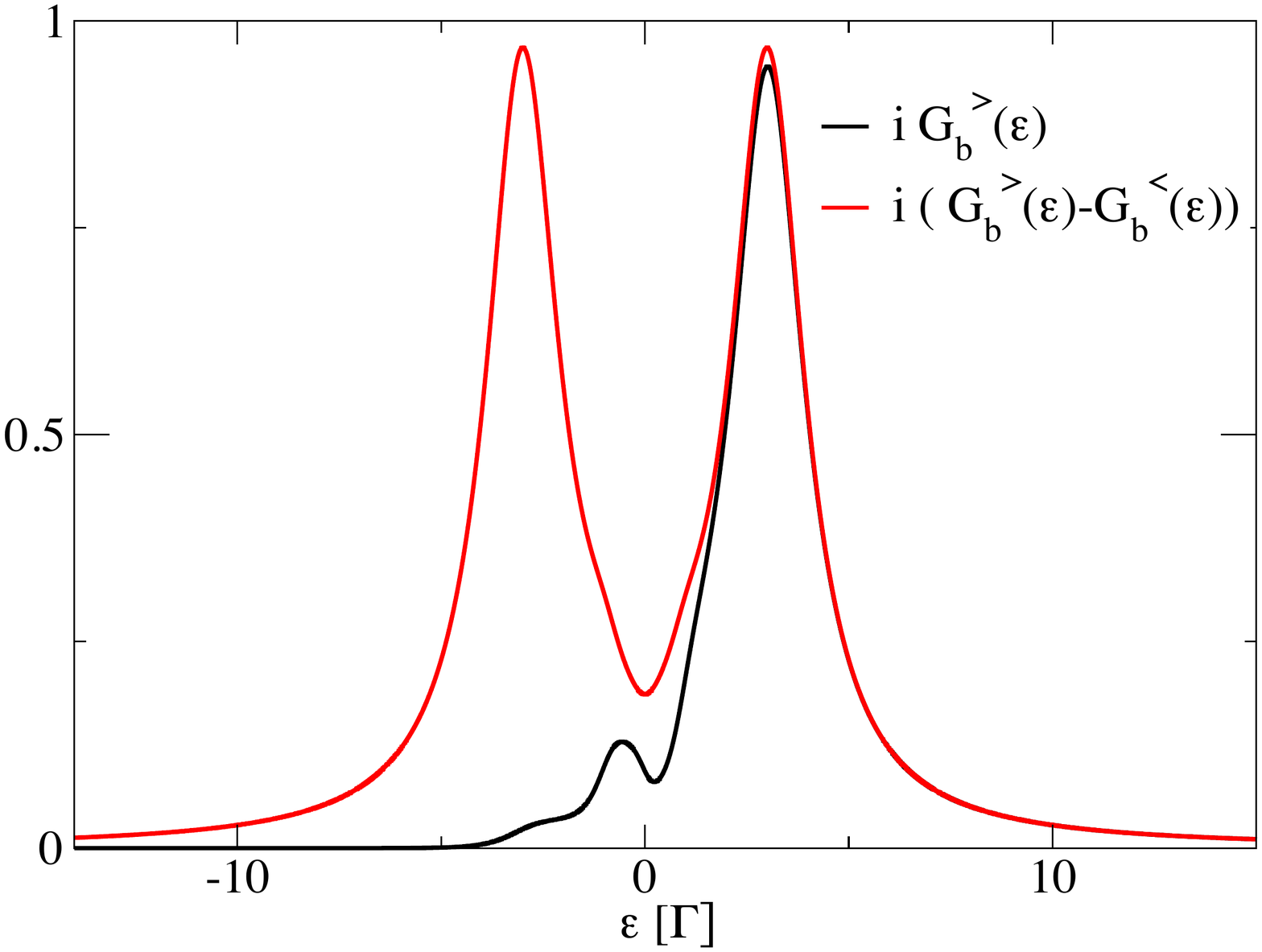}
    \end{minipage}\\
    \caption{Top panel: Charge current in $e \Gamma /\hbar$ units, evaluated in the QME and in the NCA, for $U=6 \Gamma$, $T_h=\Gamma$, and $T_c=0.2 T_h$. The renormalized energy current evaluated in the NCA, $\frac{e J^E_t}{U}$ is also presented. Bottom: Green's functions for the same parameters, and for $V$ equal to the upper panel stopping voltage.}
\label{fig2}
\end{figure}

We then consider a parameter range where sequential processes alone are not supposed to be suitable, indeed $T > \Gamma$ is not fulfilled. 
In Fig.~\ref{fig2}, the electric and renormalized energy currents are plotted as a function of renormalized bias voltage. The cotunneling processes, 
which boost the electric current at low or null voltage,
rapidly erode it for a moderate bias and reduce the stopping voltage by a factor 2.5. 
There is about a factor 2 between the NCA and the cotunneling QME current amplitudes. However, they nearly coincide concerning the stopping voltage 
($ \sim 1.83 \Gamma$ in the NCA, $\sim1.95 \Gamma$ in the QME).
The qualitative agreement between the QME and the NCA is worse for the energy current 
(up to a $U/e$ factor, red and green curves).
Indeed, interestingly, for the present parameters, there is a significant difference in the NCA between the electric drag current and the renormalized thermal current $e J^E_t /U$. 
Not only are $J^e_b$ and $J^E_t$ not proportional to each other, but also they do not even vanish for the same bias. 
At low voltage, as long as the electric current exceeds $e J^E_t /U$, the efficiency will be higher than $eV /U$. For a higher $V$, the situation is reversed.
When the drag charge current vanishes, the energy current 
corresponding to heat extracted from the hot source, is distributed -- equitably, as can be demonstrated for the symmetric case -- to the cold reservoirs, without charge current flowing through the $b$-dot. This transfer is driven by charge fluctuations in between the bottom dot and each cold reservoir separately.
For this to happen, a more elaborate structure for the Green's functions is needed, beyond the $\delta$-function shape, as occurs in the $T$-matrix QME. In the lower panel of Fig.~\ref{fig2}, the Green's functions $i G^>_b(\epsilon)$ and $i(G_b^>(\epsilon)-G_b^<(\epsilon))$ are plotted, and they reveal two broad peaks located around $\epsilon_d$ and 
$\epsilon_d+U$, as well as a low-energy finer structure in between. One has $iG_b^>(-\epsilon)=-iG_b^<(\epsilon)$ for the particle-hole
symmetric case. 

\begin{figure}[h!]
\begin{center}
\includegraphics[width=.48\textwidth]{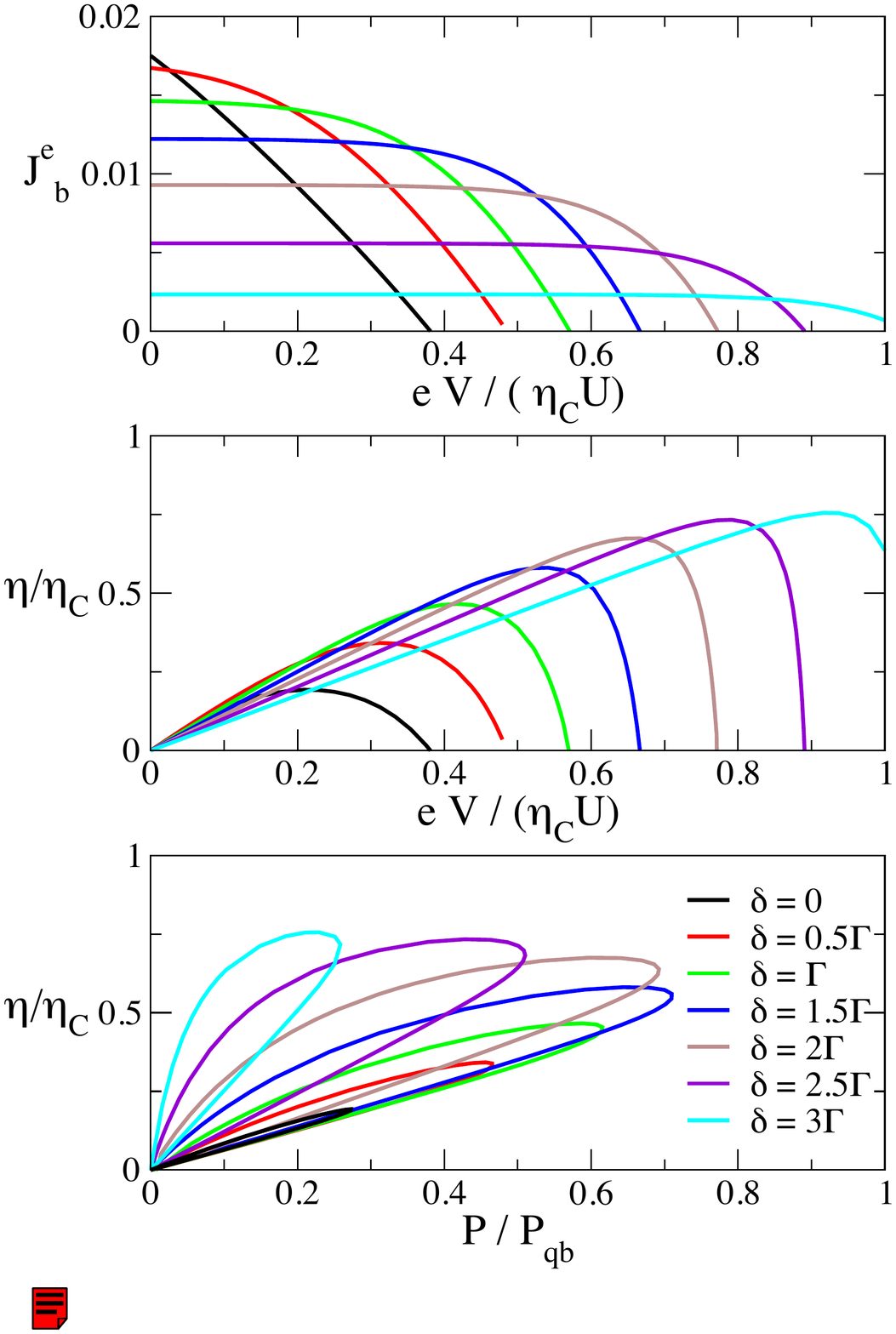}
\caption{NCA calculations for the same parameters as in Fig.~(3): $U=6 \Gamma$, $\epsilon_d =-U/2$, $T_h=\Gamma$, and $T_c=0.2 T_h$. 
The hybridization functions are shifted by $\delta$.
Top panel: electric current in $e \Gamma /\hbar$ units, middle panel: ratio of device to Carnot efficiencies as a function of bias, bottom: ratio of device to Carnot efficiencies as a function of renormalized power. See text for details.}
    \label{fig4}
\end{center}
\end{figure}

In the sequential QME, due to the $\delta$-function peaks in the spectral functions, the charge current is unaffected by a moderate bias, as seen in Fig.~\ref{fig2}. Only a bias 
roughly of the order of $U$ (it depends also on temperatures) can decrease sensibly the drag current, and as a consequence the efficiency can attain the Carnot efficiency. On the contrary, in the NCA -- as well as in the QME including cotunneling processes but for different reasons -- the voltage rapidly reduces the current, thus bounding the output power and the efficiency. 
To contain this power erosion, we consider a family of hybridization functions in Eqs.~(\ref{nosGamma}). Namely we choose shifted Heaviside 
functions
\begin{equation}
\kappa(\epsilon)= H(\epsilon-\delta) \ .
\end{equation}
For the same parameters as in Fig.~\ref{fig2}, as shown in the upper panel in Fig.~\ref{fig4}, varying $\delta$ from 0 to $\frac U 2 $ erodes the current at null bias, but it ensures a quasi-independence from
voltage in a larger range as $\delta$ grows. This can be very advantageous for efficiency and output power as revealed in the middle 
and lower panels of Fig.~\ref{fig4}. 

Despite a sensible reduction of the zero-bias current, its quasi-independence against voltage proves beneficial to maximum power and corresponding efficiency:
Going from $\delta =0$ to $\delta = 1.5\Gamma$, the first one grows by a factor 2.6, while the second one is enhanced by a factor 2.9.
These tailored hybridization functions make this device very attractive.
It is tempting to make a comparison with the maximum power that can be delivered by one channel connecting two heat sources. Whitney uncovered in Ref.~\onlinecite{Whitney14} the following upper quantum bound on output power for a two-terminal engine:
\begin{equation}
\mathcal{P}_{qb} = A_0 \frac{\pi^2}{h} k_B^2 \Delta T^2
\label{qbound}
\end{equation}
with $A_0 \simeq 0.0321$. 
Later this bound was extended to the three terminal setup~\cite{Whitney16}.
With the temperatures used in Fig.~\ref{fig4}  ($T_h=\Gamma$, $T_c=0.2 \Gamma$), this bound, Eq.~(\ref{qbound}), attains 0.0322 $\Gamma^2/\hbar$, and the lower part of Fig.~\ref{fig4} reveals that the present device achieves a maximum power of about $ 0.7 \ \mathcal{P}_{qb}$ for $\delta = 1.5 \Gamma$.

Furthermore, in the two-terminal setup, the quantum bound can be reached for a transmission function of boxcar type, for which the efficiency can be evaluated.
One finds 
\begin{equation}
\eta_{\mathcal{P}_{qb}} = \frac{1}{(\bar{T}/\Delta T)\ D+1/2}
\end{equation}
where $\bar{T}=\frac{T_h+T_c}{2}$, $\Delta T =T_h -T_c$, and $ D \simeq 2.872$. 
For the present parameters, it leads to $\eta_{\mathcal{P}_{qb}} / \eta_C \simeq 0.47$ which is surpassed in the present device. 
It appears that the large power is not achieved at the expense of efficiency. 
We can also compare the present efficiency at maximum power $\eta_{\mathcal{P}_{max}}= 0.448$ with the Curzon-Ahlborn~\cite{CA74} one for the present temperatures:
$\eta_{CA}=0.55$.

The quantum bound settled by Whitney was first obtained for a two-terminal device.
In this case, the Meir-Wingreen-Landauer exact formula [Eq.~(3) of Ref.~\onlinecite{Wingreen94}] applies with the only hypothesis of proportionate left and right lead-dot couplings, and non-interacting leads, but potentially in the presence of in-dot Coulomb interaction. Thus the output power from the interacting system is forced by this quantum bound.
To fix an order of magnitude in the two-terminal case, it can be mentioned for example, that for the problem of two orbitals in the Kondo regime (see Ref.~\onlinecite{Azema12}, Fig. 2) the maximum power per channel was about one-tenth of the quantum bound.
For a noninteracting Lorentzian transmission function, it would be even much smaller~\cite{Xining}.
 The present three-terminal setup in the Coulomb-blockade regime might not be constrained by this quantum bound which was obtained in a noninteracting theory~\cite{Whitney16}. Our numerical results do not exceed the bound but almost reach it, making the TTTE setup very attractive from an energy-harvesting perspective.

The regime $\Gamma > T_h$ would be an interesting issue to address. On the one hand, one could expect a less favorable situation than the one presented in Fig. 4, due to a likely broadening of Green's functions as $\Gamma$ increases, usually detrimental to transport. On the other hand, as in the two-terminal case, one cannot exclude for lower temperatures emerging sharp structures reminiscent of the Kondo resonance, which grows in the two-terminal device~\cite{Azema12} and is beneficial to power. Unfortunately in the case of dissymmetrical hybridizations, we encounter numerical 
difficulties~\cite{noteTemp} 
that prevent low-temperature study and
would need more numerical investment to be settled; this is left for future work.

\section{Conclusions}

We have shown that the TTTE setup presented in Fig.~\ref{Dispo} not only offers the interesting feature of decoupling 
charge and heat currents, but it also overcomes the performance of the one-dot-two-terminal setup.
It had been claimed to be "optimal"~\cite{Sanchez11} because of the possibility to attain Carnot efficiency when the
sequential QME is reliable. 
Expanding the parameter range for this device, by developing two different strategies to explore its properties, we have shown that it can also
yield an outstanding power due to its topology. 
 
The two techniques employed are very different and are successfully compared in a regime in which the sequential QME alone fails. The NCA is presumed to be valid on a wider parameter range than the QME. However, the observed qualitative agreement
-- better for the electric current --
may be reminiscent of the concordance found between the cotunneling QME and experiments in Coulomb coupled quantum dots in an unexpected parameter range~\cite{Keller16}.
 
{\it Note added.} During the review process, we became aware of some related work~\cite{Walldorf17} that addressed the issue of calculating the heat current in more general situations than the one studied here,
up to the cotunneling order.

\section*{Acknowledgments}
We gratefully acknowledge C. Mattei for detecting a few misprints.

\appendix
\section{NCA equations}

In the NCA, the true fermions are replaced by pseudo-particles~\cite{Haule01}. Assuming the following notations for fermion creation operators:
$d^\dagger_\sigma$, with $\sigma = - \bar{\sigma}=\uparrow \mathrm{or} \downarrow$, respectively, for the $b$- or $t$-dot orbital, we define the pseudofermion $f^\dagger_\sigma$ and the pseudoboson 
$a^\dagger$ and $b^\dagger$ creation operators by 
$ d_\sigma^\dagger = f_\sigma^\dagger b + \sigma a^\dagger f_{\bar{\sigma}}$. $b^\dagger$ is the bosonic operator creating the empty two-dot state, while 
$a^\dagger$ corresponds to the creation of the doubly occupied two-dot state. 

The present top and bottom fermion Green's functions can be expressed in terms of  the
pseudoparticle lesser and greater Green's functions. The relations between the Fourier-transformed Green's functions are
\begin{align}
G_{d \sigma}^<(\omega) = i \int \frac{d \omega'}{2 \pi}\Bigl[  G^<_{f \sigma}(\omega') & D^>_b(\omega'-\omega) \nonumber \\
& - G^>_{f \bar{\sigma}}(\omega'-\omega) D_a^<(\omega') \Bigr] \nonumber \\
G_{d \sigma}^>(\omega) = i \int \frac{d \omega'}{2 \pi}\Bigl[  G^>_{f \sigma}(\omega') & D^<_b(\omega'-\omega) \nonumber \\
&- G^<_{f \bar{\sigma}}(\omega'-\omega) D_a^>(\omega') \Bigr] \ .
\end{align}
The pseudoparticle Green's functions obey Keldysh equations:
\begin{eqnarray}
D^\gtrless_{a(b)}(\omega)&=& D^R_{a(b)}(\omega)  \Pi^\gtrless_{a(b)}(\omega) D^A_{a(b)}(\omega) \nonumber \\
G^\gtrless_{f \sigma}(\omega)&=& G^R_{f \sigma}(\omega)  \Sigma^\gtrless_{f \sigma}(\omega) G^A_{f \sigma}(\omega) \ ,
\end{eqnarray}
where $D^{R(A)}_{a(b)}$ are the retarded (advanced) boson Green's functions, $\Pi^\gtrless_{a(b)}$ are the boson self-energies, and
$\Sigma^\gtrless_{f \sigma}(\omega)$ are the pseudofermion self-energies.

The NCA mainly lies in the choice for the pseudoboson and pseudofermion self-energies. 
Using a diagrammatic technique, the following expressions are obtained:
\begin{eqnarray}
\Sigma_{f \uparrow}^<(\omega) =& -& \int \frac{d \omega'}{2 \pi} A_b(\omega-\omega') D_b^<(\omega') \nonumber \\
&-& \int \frac{d \omega'}{2 \pi} B_t(\omega'-\omega) D_a^<(\omega')\ ,\nonumber \\
\Sigma_{f \downarrow}^<(\omega) = &-& \int \frac{d \omega'}{2 \pi} A_t(\omega-\omega') D_b^<(\omega') \nonumber \\
&-& \int \frac{d \omega'}{2 \pi} B_b(\omega'-\omega) D_a^<(\omega')\ , \nonumber \\
\Pi^<_b (\omega) = &-& \int \frac{d \omega'}{2 \pi} B_b(\omega'-\omega) G^<_{f \uparrow} (\omega') \nonumber \\
&-&\int \frac{d \omega'}{2 \pi} B_t(\omega'-\omega) G^<_{f \downarrow} (\omega')\ , \nonumber \\
\Pi^<_a (\omega) = &-& \int \frac{d \omega'}{2 \pi} A_t(\omega-\omega') G^<_{f \uparrow} (\omega') \nonumber \\
&-&\int \frac{d \omega'}{2 \pi} A_b(\omega-\omega') G^<_{f \downarrow} (\omega') \ , 
\end{eqnarray}
whereas
\begin{eqnarray}
\Sigma_{f \uparrow}^>(\omega) = &&\int \frac{d \omega'}{2 \pi} B_b(\omega-\omega') D_b^>(\omega') \nonumber \\
&+&  \int \frac{d \omega'}{2 \pi} A_t(\omega'-\omega) D_a^>(\omega')\ , \nonumber \\
\Sigma_{f \downarrow}^>(\omega) = && \int \frac{d \omega'}{2 \pi} B_t(\omega-\omega') D_b^>(\omega') \nonumber \\
&+&  \int \frac{d \omega'}{2 \pi} A_b(\omega'-\omega) D_a^>(\omega') \ ,\nonumber \\
\Pi^>_b (\omega) = &&  \int \frac{d \omega'}{2 \pi} A_b(\omega'-\omega) G^>_{f \uparrow} (\omega')  \nonumber \\
&+& \int \frac{d \omega'}{2 \pi} A_t(\omega'-\omega) G^>_{f \downarrow} (\omega')\ ,\nonumber \\
\Pi^>_a (\omega) = && \int \frac{d \omega'}{2 \pi} B_t(\omega-\omega') G^>_{f \uparrow} (\omega') \nonumber \\
&+& \int \frac{d \omega'}{2 \pi} B_b(\omega-\omega') G^>_{f \downarrow} (\omega') \ .
\end{eqnarray}
In the preceding equations, 
\begin{eqnarray}
B_b(\omega) &=& \Gamma_{L_b}(\omega) [1-f_{L_b}(\omega)]+ \Gamma_{R_b}(\omega) [1-f_{R_b}(\omega)]\ ,\nonumber \\
B_t(\omega) &=& \Gamma_t(\omega) [1-f_t(\omega)]\ ,\nonumber \\
A_b(\omega) &=& \Gamma_{L_b}(\omega) f_{L_b}(\omega)+ \Gamma_{R_b}(\omega) f_{R_b}(\omega)\ ,\nonumber \\
A_t(\omega)&=& \Gamma_t(\omega) f_t(\omega) \ .
\end{eqnarray}
We explicitly checked, as was done also in Ref.~\onlinecite{Tosi13}, that these self-energies ensure charge and energy current conservation, $J^e_{L_b}=-J^e_{R_b}$ and $J^E_{L_b}+J^E_{R_b}+J^E_t=0$.

\section{QME Transfer rates }

In the particle-hole symmetric model $\epsilon_t=\epsilon_b\equiv\epsilon_d=-U/2$, choosing 
$\Gamma_t(\epsilon)=  \Gamma_t$, $\Gamma_{L_b}(\epsilon)=\theta(-\epsilon)\Gamma_b=\Gamma_{R_b}(-\epsilon)$, 
the sequential terms are readily evaluated.  One finds
\begin{align}
&\gamma_{0t}  =\frac 1 \hbar f_t(\epsilon_d) \Gamma_t \nonumber \\ 
&\gamma_{t0}  =\frac 1 \hbar [1-f_t(\epsilon_d)] \Gamma_t \nonumber \\
&\gamma_{0b} =\gamma_{0b}^{L_b}+\gamma_{0b}^{R_b} \ , \mathrm{with}\  \gamma_{0b}^{L_b}= \frac 1 \hbar f_{L_b}(\epsilon_d) \Gamma_b  
\  , \gamma_{0b}^{R_b}=0 \nonumber \\
&\gamma_{b0}=\gamma_{b0}^{L_b}+\gamma_{b0}^{R_b} \ , \mathrm{with}\  \gamma_{b0}^{L_b}= \frac 1 \hbar [1-f_{L_b}(\epsilon_d)] \Gamma_b \ ,\gamma_{b0}^{R_b}=0
\nonumber \\
&\gamma_{t2}=\gamma_{t2}^{L_b}+\gamma_{t2}^{R_b} \ , \mathrm{with}\ \gamma_{t2}^{L_b}=0 \ ,\gamma_{t2}^{R_b}=\frac 1 \hbar f_{R_b} (\epsilon_d+U) \Gamma_b
\nonumber \\
&\gamma_{2t}=\gamma_{2t}^{L_b}+\gamma_{2t}^{R_b} \ ,\gamma_{2t}^{L_b}=0 \ ,\gamma_{2t}^{R_b}=\frac 1 \hbar [1-f_{R_b} (\epsilon_d+U)] \Gamma_b
\nonumber\\
&\gamma_{b2}= \frac 1 \hbar f_t(\epsilon_d+U) \Gamma_t \nonumber \\
&\gamma_{2b}= \frac 1 \hbar [1-f_t(\epsilon_d+U)] \Gamma_t  \ .
\end{align}
The cotunneling terms request more calculations and a regularization procedure. 
They are of two types: some of them are local cotunneling (intradot), while the others are nonlocal cotunneling (interdot)~\cite{Jauho16}.
In the following we only consider transition rates that may affect $P_i$ and/or currents of interest.
For example, for three terminals but for an arbitrary hybridization function between dots and reservoirs, one obtains
\begin{align}
\tilde{\gamma}_{02}  =\frac 1 \hbar &\sum_{\beta_b} \int \frac{d \epsilon}{2 \pi} \Gamma_{t}(\epsilon)
\Gamma_{\beta_b}(\epsilon_t+\epsilon_b+U-\epsilon) \Bigl|\frac{1}{\epsilon-\epsilon_t+i \eta}\nonumber\\
&+\frac{1}{\epsilon_t+U-\epsilon+i \eta} \Bigr|^2
f_{t}(\epsilon)f_{\beta_b}(\epsilon_t+\epsilon_b+U-\epsilon) \ .
\end{align}
The other cotunneling terms are of the same kind. For our particular choice of $\Gamma_\alpha(\epsilon)$ and in the particle-hole symmetric case, one obtains
\begin{align}
\tilde{\gamma}_{02}  &=\tilde{\gamma}_{02}^{L_b}+\tilde{\gamma}_{02}^{R_b} \ ,\nonumber\\
\tilde{\gamma}_{20}  &=\tilde{\gamma}_{20}^{L_b}+\tilde{\gamma}_{20}^{R_b}\ , \nonumber\\
\tilde{\gamma}_{tb}  &=\tilde{\gamma}_{tb}^{L_b}+\tilde{\gamma}_{tb}^{R_b} \ ,\nonumber\\
\tilde{\gamma}_{bt}  &=\tilde{\gamma}_{bt}^{L_b}+\tilde{\gamma}_{bt}^{R_b}  \ ,
\end{align}
with
\begin{eqnarray}
\tilde{\gamma}_{02}^{L_b}= \tilde{\gamma}_{20}^{R_b}=\tilde{\gamma}_{tb}^{L_b}=\tilde{\gamma}_{bt}^{R_b} =\tilde{R}_1+I_1 \ , \nonumber \\
\tilde{\gamma}_{02}^{R_b}= \tilde{\gamma}_{20}^{L_b}=\tilde{\gamma}_{tb}^{R_b}=\tilde{\gamma}_{bt}^{L_b} =\tilde{R}_2+I_2 \ .
\end{eqnarray}
In the preceding expressions,
\begin{align}
\tilde{R}_1 =& \frac{\Gamma_b \Gamma_t}{\hbar} \int_0^{+\infty} \frac{d \epsilon}{2 \pi}\frac{1}{(\epsilon -\epsilon_d)^2} f_t (\epsilon) [1-f_{R_b}(\epsilon)] \ ,\nonumber\\
\tilde{R}_2 =& \frac{\Gamma_b \Gamma_t}{\hbar} \int_0^{+\infty} \frac{d \epsilon}{2 \pi}\frac{1}{(\epsilon -\epsilon_d)^2} f_{R_b} (\epsilon) [1-f_{t}(\epsilon)] \ ,\nonumber\\
I_1 =&  \frac{\Gamma_b \Gamma_t}{\hbar} \int_0^{+\infty} \frac{d \epsilon}{2 \pi} \frac{1}{(\epsilon +\epsilon_d)^2+\eta^2} \Bigl[ 1-2\frac{\epsilon+\epsilon_d}{\epsilon-\epsilon_d} \Bigr]
f_t (\epsilon) [1-f_{R_b}(\epsilon)] \ ,\nonumber\\
I_2 =&\frac{\Gamma_b \Gamma_t}{\hbar} \int_0^{+\infty} \frac{d \epsilon}{2 \pi} \frac{1}{(\epsilon +\epsilon_d)^2+\eta^2} \Bigl[ 1-2\frac{\epsilon+\epsilon_d}{\epsilon-\epsilon_d} \Bigr]
f_{R_b} (\epsilon) [1-f_{t}(\epsilon)] \ .
\end{align}
$\tilde{R}_1$ and $\tilde{R}_2$ are well-behaved expressions, whereas $I_1$ and $I_2$ diverge, and one needs to extract manually these diverging terms.
The divergence that needs to be cured is inherent to the present $T$-matrix-based QME method~\cite{Koller10}. 
In the current calculation, one needs the difference $I_1-I_2$, which can be written
\begin{equation}
I_1-I_2 = \int_{\epsilon_d}^{+\infty} d \epsilon \frac{F(\epsilon)}{\epsilon^2+\eta^2}
\end{equation}
with
\begin{equation}
F(\epsilon)= \frac{\Gamma_t \Gamma_b}{2\pi \hbar} \Bigl( 1 - 2 \frac{\epsilon}{\epsilon-2 \epsilon_d} \Bigr) 
\Bigl( f_t(\epsilon-\epsilon_d) - f_{R_b}(\epsilon-\epsilon_d)  \Bigr)  \ .
\end{equation}
We cannot proceed analytically further as was done, for example, in Refs.~\onlinecite{Jauho16,Lim17}, because in the preceding equation the two Fermi functions are characterized by different temperatures. 
We thus proceed along the lines of Ref.~\onlinecite{Turek02}, throwing away the divergent part, which is in fact a sequential 
contribution, by making the substitution
\begin{equation}
I_1-I_2 \rightarrow \mathcal{P} \int_{\epsilon_d}^{+\infty} d \epsilon \frac{F(\epsilon)-F(0)}{\epsilon^2} \ ,
\end{equation} 
where $\mathcal{P}$ indicates principal part.
Finally the current expression requires the following term $\tilde{R}_1-\tilde{R}_2 +I_1-I_2$, which reads after regularization
\begin{eqnarray}
\Delta = \frac{\Gamma_b \Gamma_t}{\hbar} \int_0^{+\infty} \frac{d \epsilon}{2 \pi}\frac{1}{(\epsilon -\epsilon_d)^2} ( f_t (\epsilon) -f_{R_b}(\epsilon))
\nonumber\\
+ \mathcal{P} \int_{\epsilon_d}^{+\infty} d \epsilon \frac{F(\epsilon)-F(0)}{\epsilon^2} \ .
\end{eqnarray}


\begin{thebibliography}{99}
\bibitem{noteTemp} Low temperatures  ($T_h \ll \Gamma$) cannot be reached in the case of nonproportionate hybridization functions within our algorithm: the self-consistent NCA equations do not converge.
\end{thebibliography}
\end{document}